# Differences of interface and bulk transport properties in polymer field-effect devices

S. Grecu, M. Roggenbuck, A. Opitz, W. Brütting[*]

*Experimentalphysik IV, Universität Augsburg, 86135 Augsburg, Germany*

**Abstract**

The influence of substrate treatment with self-assembled monolayers and thermal annealing was analysed by electrical and structural measurements on field-effect transistors (FETs) and metal-insulator-semiconductor (MIS) diodes using poly(3-hexylthiophene) (P3HT) as a semiconducting polymer and Si/SiO$_2$ wafers as a substrate.

It is found that surface treatment using silanising agents like hexamethyldisilazane (HMDS) and octadecyltrichlorosilane (OTS) can increase the field-effect mobility by up to a factor of 50, reaching values in saturation of more than $4\times10^{-2}$ cm$^2$/Vs at room temperature. While there is a clear correlation between the obtained field-effect mobility and the contact angle of water on the treated substrates, X-ray diffraction and capacitance measurements on MIS diodes show that structural and electrical properties in the bulk of the P3HT films are not influenced by the surface treatment. On the other hand, thermal annealing is found to cause an increase of grain size, bulk relaxation frequency and thereby of the mobility perpendicular to the SiO$_2$/P3HT interface, but has very little influence on the field-effect mobility. Temperature dependent investigations on MIS diodes and FETs show that the transport perpendicular to the substrate plane is thermally activated and can be described by hopping in a Gaussian density of states, whereas the field-effect mobility in the substrate plane is almost temperature independent over a wide range. Thus, our investigations reveal significant differences between interface and bulk transport properties in polymer field-effect devices.

*PACS:* 61.41.+e ; 61.10.Eq; 72.80.Le; 73.61.Ph; 82.35.Gh

*Keywords:* Organic transistor; MIS diode; Self-assembled monolayer; Polythiophene; P3HT

---

[*] Corresponding author. *E-mail address*: Wolfgang.Bruetting@physik.uni-augsburg.de

# 1. Introduction

In recent years conjugated polymers have gained increasing interest as active materials in organic electronics. Their wide range of chemical variability in combination with low-cost solution processing make them attractive candidates for the fabrication of electronic devices such as light-emitting diodes (LEDs), photovoltaic cells (PVCs) and field-effect transistors (FETs)[1,2]. For the latter, high field-effect mobility is required and this is usually associated with high degree of structural order of the films. Poly(alkylthiophenes) have proven to be very promising candidates to achieve this goal. Charge carrier mobility as high as 0.1 cm$^2$/Vs has been reported in regioregular poly(3-hexylthiophene) (rr-P3HT)[3], its structure being shown in Fig. 1.

In field-effect transistors the charge transport in the active channel is restricted to a few monolayers close to the interface[4], so it is to be expected that the surface treatment of the underlying substrate has a strong influence on the charge carrier mobility of P3HT FETs[5]. Another issue that comes into play is the anisotropy of charge carrier transport in polymeric field-effect devices. The regular arrangement of the alkane side groups promotes the formation of a lamellar stacking of the polymer chains with good π-orbital overlap between neighbouring chains. This orientation is expected to result in anisotropy of the mobility in the parallel and perpendicular direction with respect to the substrate. From the comparison of MIS diodes and FET data Scheinert *et al.*[6,7] and our group[8] have suggested that the charge carrier mobility in the substrate plane and perpendicular to it may differ by up to four orders of magnitude. Contrarily, Tanase *et al.*[9] have shown that differences between the values of the field-effect mobility and the values obtained from space-charge limited currents in diode structures is due to a charge carrier density dependence of the mobility in disordered hopping systems. In amorphous films of OC$_1$C$_{10}$-PPV and P3HT having field-effect mobility in the range $10^{-5}$-$10^{-4}$ cm$^2$/Vs, at room temperature, no evidence for a pronounced anisotropy of charge carrier transport was found.

In this paper we will demonstrate that by suitable substrate treatment the field-effect mobility of P3HT can be varied over a wide range whereby the surface energy of the Si/SiO$_2$ substrate will turn out to be the decisive control parameter. We will further show that by silanisation highly ordered films with field-effect mobility approaching 0.1 cm$^2$/Vs can be achieved in which the mobility in the substrate plane and perpendicular to it is strongly anisotropic.

# 2. Experimental

## 2.1. Sample preparation

The substrates used in this work were highly p-doped Si wafers (1-5 mΩ cm) with high quality, thermally grown oxide as the gate insulator. Four different types of surface treatments were em-ployed prior to the deposition of P3HT. The first substrate type was only wet-cleaned in an ultrasonic bath with solvents (acetone, isopropyl), in the following termed untreated. The other three were additionally exposed to oxygen plasma in order to create a hydrophilic surface which is required for the growth of self assembled monolayers through a silanisation reaction. One of these substrates, to which we will refer to as the $O_2$ plasma treated one, was used to prepare field-effect devices without any further treatment. The other two were used for silanisation immediately after the $O_2$ plasma treatment. Two types of silane molecules, hexamethyldisilazane (HMDS) and octadecyltrichlorosilane (OTS), respectively, were used. The HMDS treatment was done in the liquid phase at 60°C for 24 hours. The OTS treatment was carried out at room temperature, in a solution of OTS in n-heptane (0.86 mM), in an exsiccator. Samples were subsequently cleaned from residual OTS in chloroform in an ultrasonic bath. The values of the contact angles of water droplets on these substrates are summarized in Table 1.

The MIS diodes have been prepared on Si substrates with a 50 nm thin oxide layer (Fig. 1(a)). Solutions of regioregular P3HT in toluene (5 %wt.) were spin-coated at 2000 rpm resulting in films with a thickness of about 240-265 nm, except for the $O_2$ sample where it was about 130 nm. The P3HT was obtained from Merck and used without any additional purification. The molecular weight of P3HT was about 14,000. After a drying step (12 h, $10^{-3}$ mbar, 330 K), for all samples, top gold contacts were evaporated through a shadow mask, defining an area of about 28-33 mm$^2$. The FETs were prepared by spin-coating a 50 nm thick P3HT film onto Si/SiO$_2$ substrates with photolithographically patterned source and drain electrodes from Au, having a circular geometry to reduce leakage currents. The insulator thickness in this case was 200 nm, the channel length 5 μm and channel width 1000 μm. Thus, the resulting FETs were in the bottom-gate, bottom-contact configuration (Fig. 1). To avoid unintentional doping by oxygen the whole film preparation was performed in a glove-box system under inert atmosphere (<0.1 ppm oxygen and <1 ppm water content). The substrates for the X-Ray diffraction analysis were normal glass. In this case the P3HT films were spin-coated from a 5 %wt. solution in toluene at 1000rpm resulting in a thickness of about 400-450 nm. Additionally, heating at 350 to 420 K in high vacuum was applied to the samples in order to study the influence of further thermal annealing.

*2.2. Measurement details*

The X-ray diffraction (XRD) measurements were performed on a Siemens D-5000 diffractometer in grazing incidence geometry (0.03° between the incident beam and the sample, wavelength 1.54 Å), on pristine and annealed films for all types of samples.

The MIS diodes were characterized by measuring the capacitance–voltage (C–V) and capacitance–frequency (C–f) dependence, using a Solartron 1260 impedance/gain-phase analyzer coupled with a Solartron 1296 dielectric interface, in the frequency range from 1 to $10^6$ Hz. An a.c. amplitude of 0.5 V and bias between −10 V and +10 V was used to operate the diodes either in accumulation or depletion. Measurements of the output and transfer characteristics of FETs were performed using two independent source-measure units (Keithley 236). Both types of devices were measured and annealed in a cryostat in high vacuum (<$10^{-6}$ mbar) and darkness. The transfer between the glove-box and the cryostat was performed in a load-lock system so that the samples were not exposed to air at any time.

## 3. Results and discussion

### 3.1. Structural investigations

The structural ordering of the P3HT films was analysed by XRD measurements. P3HT films were prepared on untreated as well as $O_2$ plasma, HMDS and OTS treated substrates, as previously mentioned. Additionally the changes upon thermal annealing of the films (up to 24 h, 350 K) were analysed. A typical XRD spectrum of a P3HT film deposited on an OTS treated substrate is shown in Fig. 2(a). Clearly visible is the (100) peak at about 5.5° and additionally, much weaker, the (200) peak at 11°. The effect of thermal annealing resulting in an increase of peak height is shown as an example for the untreated substrate in Fig. 2(b).

The observed first order diffraction peaks were fitted in order to calculate the lattice constant $a$ from the Bragg condition

$$2a\sin(\theta) = n\lambda \tag{1}$$

In polycrystalline materials, the crystallite size $l$ can be obtained from the Scherrer equation using the relation between full width at half maximum $FWHM$ ($2\theta$) of the diffraction peak and the diffraction angle $2\theta$

$$FWHM(2\theta) = 0.94 \cdot \lambda / l \cdot \cos(\theta). \tag{2}$$

In polymers, however, the FWHM is usually dominated by variations in interplanar spacings. Thus the width of the XRD peaks can only give a qualitative hint to the degree of crystallinity. The calculated crystallite size from the Scherrer equation is about 10-11nm for the as-prepared samples and increases up to 13 nm upon annealing. The increase of peak intensity after annealing indicates an increase of crystallinity which can be due to a higher number of crystallites and also an increase of crystallite size. As other groups have shown, the structure model of rr-P3HT films consists of crystallites embedded into an amorphous polymer matrix[10,11]. All spectra show a diffraction peak at about 5.4° which is known for the organized lamellar structure of rr-P3HT with π-π-interchain stacking within the crystallites as shown in the inset of Fig. 2(a). This corresponds to a layer spacing for the (100) direction of about $a$=16 Å which is comparable to literature data[12]. No evidence for the other two possible orientations of P3HT with respect to the substrate (corresponding to diffraction peaks around 23°) was found.

*3.2. Capacitance measurements on MIS diodes*

Capacitance-voltage (*C-V*) and capacitance-frequency (*C-f*) measurements were performed on MIS diodes prepared with different substrate treatment (Figs. 3 and 4). In general, the *C-V* characteristics of p-conducting organic MIS diodes show two regimes: accumulation at negative bias and depletion at positive bias with the measured value of *C* corresponding either to the insulator capacitance ($C_{ins}$) or to the series sum of the insulator capacitance and the capacitance of the organic semiconductor (($C_{ins}^{-1}+C_S^{-1})^{-1}$), respectively. The absence of inversion is explained by the extraordinarily long generation times for minority carriers in wide-gap organic semiconductors[7].

Using the standard Schottky–Mott analysis, the doping concentration $N_A$ can be extracted from the slope of the *C–V* curves in the transition region between accumulation and depletion via[13]

$$\frac{\partial}{\partial V}\left(C^{-2}\right)=\frac{2}{\varepsilon_0\varepsilon_S q N_A A^2}, \tag{3}$$

where $\varepsilon_0$ is the permittivity of vacuum, $\varepsilon_S$ the dielectric constant of the semiconductor and *A* the active diode area. Thus $N_A$ can be extracted from a plot of $1/C^2$ vs. the voltage applied to the Si bottom gate electrode as shown in Fig. 3(b) for the untreated sample. Also included in this figure is a simulation using an analytical expression for the *C-V* characteristics given by Scheinert and Paasch[7]. Within the error limits, both methods give the same values for the doping concentration $N_A$. We note that the simulated curves show inversion at voltages above 2 V, which is not seen experimentally for reasons discussed above. The simulation also yields information on the flat-band voltage. In an MIS structure the flat-band voltage

$$V_{FB} = \phi_{MS} - \frac{Q_{SS}}{C_{ins}} \tag{4}$$

incorporates the metal-semiconductor work function difference $\phi_{MS}$ and the equivalent fixed oxide charges $Q_{SS}$[14].

Applying the Schottky-Mott analysis to the *C-V* curves shown in Fig. 3(a) yields very similar doping concentrations in the range 2.5-7×10$^{15}$ cm$^{-3}$ for all samples, with the exception of the O$_2$ plasma treated one, where $N_A$ is up to one order of magnitude higher (see Table 2). A closer look at the *C-V* curves reveals some interesting details: Whereas the untreated, HMDS and OTS-treated MIS diodes show a sharp transition between accumulation and depletion at zero or slightly positive flat-band voltage with only marginal hysteresis between increasing and decreasing bias sweeps, the O$_2$ plasma treated diode has a larger hysteretic behaviour and the transition between accumulation and depletion takes place at higher $V_{FB}$ (between 2.1 and 3.6 V). This can be seen as an indication of interface states at the P3HT/SiO$_2$ interface created by the O$_2$ plasma treatment. We note that the measured capacitance does not saturate at the value of the oxide capacitance in the accumulation regime, but increases significantly above $C_{ins} \approx 70$ nF/cm$^2$. This can be explained by charge spreading beyond the active area defined by the top contact[15]. This effect is particularly pronounced for the OTS treated substrate which can be seen as an indication for the high in-plane charge carrier mobility for this treatment (see below).

The *C-f* curves measured at negative gate bias show a relaxation step from the accumulation capacitance to the lower value in depletion (Fig. 4(a)). The reason is the finite semiconductor bulk conductivity which sets an upper limit for the frequency up to which the injected majority carriers can follow the applied a.c. frequency and reach the accumulation layer at the interface to the oxide. Thus, this relaxation process yields information on the transport properties perpendicular to the substrate. The relaxation frequency $f_R$ can be obtained from the maximum in the dielectric loss function (the conductance $G$ divided by the angular frequency ω) as shown in Fig. 4(b). Using a simple equivalent circuit (with a capacitance $C_{ins}$ for the insulator, one RC element for semiconductor and a lead resistance $R_L$, see Fig. 4(c)) the corresponding relaxation time $\tau_R = (2\pi f_R)^{-1}$ then directly yields the semiconductor bulk resistance according to[8,16]

$$\tau_R = R_S (C_{ins} + C_S). \qquad (5)$$

We note that the simple circuit used here is not sufficient for a full description of the dielectric response of these MIS structures. For example, it has been demonstrated by Scheinert and Paasch[7] that another *RC* element for the accumulation layer should be included in the circuit. A comparison of fits using both the simple and the extended circuit model is included in Fig. 4(b). Even more sophisticated models have been suggested recently by Torres *et al.*[16]. However, by comparing fits with different circuit models we have carefully checked that the error in the semiconductor bulk resistance using the simplified analysis used here is always less than 20%.

With the knowledge of the doping concentration $N_A$ from the *C-V* analysis one can then calculate the charge carrier mobility $\mu_\perp$ perpendicular to the insulator via:

$$R_S = \frac{d_S}{qN_A \mu_\perp A}. \qquad (6)$$

From the comparison of different samples as shown in Fig. 4 one can notice two important features: On the one hand, substrate treatment has only a weak influence on the relaxation frequency, but on the other hand there is a strong effect of annealing on it. Whereas the pristine MIS diodes show relaxation frequencies in the range of 10 to 100 Hz the annealed devices have significantly higher values of $f_R$ in the kHz range (see Fig. 4(b) and Table 2). This clearly indicates that the semiconductor bulk resistance in MIS diodes can be drastically reduced by up to three orders of magnitude by thermal annealing, whereas substrate treatment has little influence on it. Since the changes in doping upon annealing as determined from the *C-V* data are comparatively small, one has to conclude that the charge carrier mobility perpendicular to the insulator interface is strongly enhanced by annealing and can be correlated with the increase of crystallinity, as seen in the XRD measurements.

*3.3. Current-voltage characteristics of OFETs*

The output and transfer characteristics of bottom-gate bottom-contact FETs for the different treatments of $SiO_2$ substrates are shown in Fig. 5. The output characteristics (Fig. 5a) show a linear onset with little evidence for contact resistances and a well-defined saturation region for $|V_D| > |V_G|$. As seen from the output characteristics, at a given gate voltage (−10V), the current increases, from the untreated substrate towards the OTS-treated one, by more than one order of magnitude. This can either be caused by an increase of the charge carrier mobility or/and by a positive shift of the threshold voltage. Therefore one has to look at the transfer characteristics in more detail.

In the saturation regime the drain current is given by:

$$I_{D,Sat} = \frac{W}{2L} \mu \cdot C_{ins} (V_G - V_T)^2. \tag{7}$$

Therein *W* and *L* are the channel width and channel length, respectively, $C_{ins}$ is the specific capacitance of the oxide, $V_G$ is the gate voltage and $V_T$ the threshold voltage. The threshold voltage $V_T$ in the Shockley model is related to the flat-band voltage, the bulk doping $N_A$ and the bulk potential $\phi_B$ and is classically defined as the voltage where inversion starts:

$$V_T = V_{FB} + \frac{eN_A d_S}{C_{ins}} - 2\varphi_B. \tag{8}$$

Here the transistors operate only in accumulation and the threshold voltage is determined as an experimental parameter from the fit of $\sqrt{I_{D,Sat}}$ vs. $V_G$ (Fig. 5(c)). Additionally, for analysing the subthreshold behaviour the switch-on voltage $V_{SO}$ and the subthreshold slope were determined. The former is defined as the gate voltage where the current starts to increase[17] in the semi-logarithmic plot (Fig. 5(b)). The resulting parameters for the four different substrate treatments before and after annealing are summarized in Table 3.

Comparing the different treatments the most striking observation is the increase of the field-effect mobility by a factor of about 40 from the $O_2$ plasma treated substrate to the OTS-treated one before annealing (after annealing the increase is approximately 50). At the same time, the FET on the untreated substrate has the highest ON-OFF ratio, the most negative threshold voltage, and the lowest value of the inverse subthreshold slope. The ON-OFF ratio for the OTS-treatment is significantly lower but still acceptable. The HMDS-treated substrate has a high threshold voltage and also a high value of the inverse subthreshold slope. The highest value of $V_T$, however, is found for the $O_2$ plasma treated sample.

After thermal annealing the transfer characteristics show a shift of the switch-on and threshold voltage, but the magnitude and direction of the shift are not the same for all the samples. The FET on the untreated substrate shows a positive shift, whereas the shift for the OTS treated sample is smaller and negative. The HMDS treated FET is almost unchanged, while the $V_T$ of the $O_2$ plasma treated sample increases dramatically. Further, the ON-OFF ratio and the inverse subthreshold slope are slightly improved by thermal annealing. Most remarkable, however, is the fact that there is no significant change in the field-effect mobility, reflected by the slopes of the curves in Fig. 5(c).

A comparison to MIS diode parameters (Table 2) does not show a direct correlation between flat-band voltage and threshold voltage as would be expected from Eq. (8). Nevertheless, there is a clear indication that oxygen plasma treatment leads in both devices to high positive values of $V_{FB}$ and $V_T$.

*3.4. Temperature dependence of MIS diode and FET characteristics*

In order to get further insight into the transport mechanism in the substrate plane and perpendicular to it, we measured the temperature dependence of the *C-f* characteristics in MIS diodes and the transfer characteristics in FETs. The obtained curves in the temperature range between 200 and 420 K are shown in Fig. 6 for the OTS treated substrates. As shown in Fig. 6(a), the bulk relaxation frequency in the MIS diode shifts over almost four orders of magnitude in the investigated temperature range. We have also measured the temperature dependence of the *C-V* characteristics (not shown here). As reported in the literature[7], the flat band voltage is shifting towards higher positive values with increasing temperature; however, no significant changes in the doping concentration with temperature were observed up to 350 K. Thus, the strong shift of $f_R$ directly reflects a strong temperature dependence of the perpendicular mobility in MIS diodes. Contrarily, in FETs (see Fig. 6(b)) the current at large gate voltage changes by only a factor of two over the whole temperature range indicating a very weak temperature dependence of the field-effect mobility parallel to the substrate. Even if the temperature is lowered to 40 K (not shown here) the mobility stays at about $1\times10^{-2}$ cm$^2$/Vs.

In Figure 7(a) the temperature dependence of the mobility for MIS diodes and field-effect transistors on the different substrates is shown. The change of the mobility in MIS diodes amounts to several orders of magnitude for all the samples and shows the typical behaviour of hopping transport.

It has been shown that charge transport in semiconducting polymers is well described by hopping models based on a Gaussian density of states[18,19,20,21]. Considering the correlated disorder model, where the expression for the mobility at low electric fields is given by

$$\mu(E \to 0) = \mu_0 \exp\left[-\left(\frac{3\sigma}{5kT}\right)^2\right] \qquad (9)$$

with $\sigma$ the width of the Gaussian DOS, $k$ Boltzmann´s constant, $T$ absolute temperature and $\mu_0$ the microscopic mobility. Values for $\sigma$ of 98 meV for the OTS treated MIS diode, 118 meV for the HMDS treated one, 110 meV for the $O_2$ treated one and 105 meV for the untreated substrate are obtained, while the prefactor $\mu_0$ lies in the range 0.2-4×10$^{-3}$ cm$^2$/Vs for all samples. As $\sigma$ is a measure of disorder, this indicates that the degree of disorder in the bulk of the polymer films increases from the OTS treated substrate to the HMDS treated ones. The observed decrease of mobility above 380 K for the untreated, HMDS and OTS treated substrates is probably due to morphological changes rather than thermal degradation of the polymer. The $O_2$ treated sample, however, shows a steep increase of the perpendicular mobility at temperatures above 335 K, which could be related to changes at the oxygen plasma treated SiO$_2$-P3HT interface upon annealing at higher temperatures. This indicates that this substrate treatment is less stable as compared to the other three methods.

In contrast to the MIS diodes, the temperature variation of the field-effect mobility is very small. Among the different substrates, the OTS treated one shows the weakest dependence which also results in the highest mobility values at all temperatures. For the other samples, with smaller values of the field-effect mobility, the temperature dependence is somewhat stronger, but still less than one order of magnitude.

A central issue with respect to the observed big differences in the mobilities between FETs and MIS diodes is the question whether this is caused by a density dependent charge carrier mobility. Therefore we have compared our data with the predictions of an isotropic hopping model developed recently by Pasveer et al.[20,21]. As an example, Fig. 7(b) shows a comparison of the temperature dependent mobility data for OTS treated structures with simulations based on this model. The required parameters are a hopping distance of 1.4 nm and a disorder parameter $\sigma$=100 meV which are in going with the numbers given in Ref. 9. The prefactor for the hopping mobility is adjusted to give the best agreement with our data for the MIS diode. Since there are no other free parameters in this model, the only difference between MIS diodes and FETs lies in the charge carrier densities. For the simulation we have taken the value of the doping $N_A = 5×10^{15}$ cm$^{-3}$ for the MIS diode and the esti-mated density of field-effect induced carriers at $V_G = -20$ V of about $5\times10^{19}$ cm$^{-3}$ (in agreement with Ref. 9). The comparison shows that the different temperature dependencies of the mobility in MIS diodes and FETs are qualitatively reproduced by the isotropic hopping model; however, there is a quantitative difference of about two orders of magnitude by which the experimentally observed FET mobility is higher than predicted by this model.

*3.5. Discussion*

In this work, we have addressed interface and bulk transport in polymeric field-effect devices. We have seen significant differences between the field-effect mobility parallel to the polymer-SiO$_2$ interface and the mobility perpendicular to it. The room temperature mobility values in MIS diodes and FETs for different treatments are listed in tables 2 and 3. The values are distinguished by surface treatment, which is characterized by a distinct contact angle, as well as by heat treatment. Clearly, the field-effect mobility increases by almost two orders of magnitude with the surface treatment, but does not change very much due to thermal annealing over the measured temperature range. Contrarily, the perpendicular mobility in MIS diodes increases upon annealing by up to four orders of magnitude and becomes almost independent of the substrate treatment. Thus, the effect of substrate treatment and thermal annealing is complementary in FETs and MIS diodes. Interestingly, the field-effect mobility shows a close correlation with the water contact angle on the substrate, increasing with the hydrophobicity of the SiO$_2$ surface. Such behaviour has recently also been observed by Veres *et al.*[5]. Fig. 8 shows their data together with the values obtained in this work. The excellent agreement is remarkable; the only exception being the O$_2$ plasma treated FET. However, this may be related to the fact that the O$_2$ plasma treatment additionally causes acceptor-like doping of the P3HT layer, which can enhance the observed mobility[22]. Thus one has to conclude that the modification of the surface energy of silicon oxide by appropriate treatments is of outmost importance for achieving high field-effect mobilities in OFETs with P3HT as a semiconducting polymer. As an explanation one can take the suggested lamellar structure of the P3HT chains (see. Fig. 2) in which the aliphatic side chains are standing upright on the substrate. Obviously, this structural motive within the first few monolayers is crucial for achieving good π-π-interchain stacking and thus high charge carrier mobility parallel to the surface of the SiO$_2$ gate dielectric.

These results are in going with recent publications on the charge carrier distribution within the active channel of organic FETs. Tanase *et al*. have shown that in disordered organic FETs the charge carrier density decreases already by one order of magnitude within the first nanometers away from the gate dielectric[4]. Taking into account the carrier density dependence of the mobility in disordered systems they could show that the macroscopically measured field-effect mobility corresponds to the local mobility of charge carriers directly at the interface. By performing in-situ electrical measurements during layer growth Dinelli *et al.* came to a similar result for FETs with ordered layers of sexithienyl molecules[23]. They could show that the first two monolayers next to the dielectric interface dominate the charge transport. Therefore, it is understandable that the substrate treatment in bottom-gate FETs, based on rr-P3HT, has such a strong influence on the field-effect mobility. If this crucial parameter is not properly controlled in the experiment, large scattering of the field-effect mobility can be obtained in nominally identical devices (see e.g. the compilation of data from different groups in Ref. 5). Further consequences of different surface treatments are seen in variations of related transistor parameters such as switch-on or threshold voltage, ON-OFF ratio and sub-threshold slope. Pernstich *et al.* have recently published a comprehensive survey of these effects in pentacene FETs[24]. Although we have not studied these effects in detail, our data show that substrate treatment also has a strong influence on these parameters. Their control will be crucial for fabricating electronic circuits based on organic FETs.

Another outcome of our work is the huge difference of charge carrier mobility for OFETs and MIS-diodes. The perpendicular mobility is orders of magnitude smaller than the field-effect mobility parallel to the interface even after thermal annealing. Such differences between in-plane and sandwich geometry have been observed before on polymeric devices[6-9]. However, there was a debate whether this difference is due to a structural and electrical anisotropy of the polymer film or a consequence of the mobility depending on the carrier density. As Tanase *et al.*[9] have demonstrated by a comparison of FET mobility data with mobility data obtained from space-charge limited currents on diodes an apparent anisotropy of about a factor of 10 could be explained in amorphous polymers by the latter effect. Nevertheless, they also found that in ordered polymers (e.g. $OC_{10}C_{10}$-PPV) there exists a true electrical anisotropy which is not explained by different carrier densities in OFETs and diodes[25]. For the case of P3HT, they have observed FET mobilities up to $6\times10^{-4}$ cm$^2$/Vs at $V_G = -19$ V which is almost two orders of magnitude lower than the mobilities obtained here on silanised substrates. Thus it is questionable whether P3HT as prepared here can be treated as an amorphous polymer. The XRD data rather indicate that our P3HT films are nanocrystalline. Such a picture has recently been put forward also by Street *et al.* for regioregular poly(3,3'''-didodecyl-quarterthiophene) (PQT-12)[26]. We therefore suggest that the observed huge differences between the mobilities for transport parallel and perpendicular to the gate insulator interface are caused by a structural anisotropy. Preliminary ellipsometric measurements on our P3HT films have also indicated considerable optical anisotropy in P3HT comparable to poly(3-octylthiophene)[27].

Moreover, the most convincing argument for a true electrical anisotropy of P3HT films comes from our temperature dependent investigations on MIS diodes and FETs. The temperature dependent mobility data for both types of devices clearly show that charge carrier transport parallel and per-pendicular to the dielectric interface can not be described by an isotropic hopping model. Thus, highly ordered films of rr-P3HT display a true electrical anisotropy which emphasises the importance of the first few monolayers of the film for charge transport even more.

## 4. Conclusion

In conclusion, we have found that the mobility in P3HT field-effect transistors depends very sensitively on the nature of the interface to the gate dielectric. Using silanisation of $SiO_2$ with different agents the mobility can be increased by almost two orders of magnitude and reaches values of about $4\times10^{-2}$ cm$^2$/Vs. Interestingly, the structural ordering in the bulk of the films as controlled by thermal annealing has almost no influence on the field-effect mobility.

By comparing OFETs with MIS diodes we have found a large difference of charge carrier mobility parallel and perpendicular to the gate dielectric which can not be explained alone on the basis of a carrier density dependent hopping mobility. Owing to their lamellar structure, highly ordered P3HT films thus display anisotropic transport which needs to be considered in extended hopping models.

## 5. Acknowledgment

This work was supported by the Deutsche Forschungsgemeinschaft (Focus Programme 1121, "Organic Field-Effect Transistors") and the German Bundesministerium für Bildung und Forschung (Focus Programme "Polymer Electronics", POLITAG project). We thank Dago de Leeuw and Maxim Skhunov for helpful discussions and the companies Merck and Philips for providing materials and financial support.

# Figure captions

**Fig. 1:** (a) Device structure of MIS diode and field-effect transistor. (b) Chemical structure of regioregular poly(3-hexylthiophene).

**Fig. 2:** Gracing incidence x-ray diffraction patterns of P3HT films on glass substrate. (a) P3HT on OTS treated substrate with first and second order diffraction peak. The inset shows a cartoon of the lamellar structure and layer spacing. (b) XRD peak of P3HT on an untreated substrate before and after annealing for 12 h at 350 K.

**Fig. 3:** (a) Capacitance-voltage characteristics of Au/P3HT/SiO$_2$/Si MIS diodes. The curves for the untreated, HMDS and OTS treated SiO$_2$ surface are shown only after annealing because there are no significant changes during annealing (measurement frequency = 1Hz, apart from the untreated one, measured at 10Hz). The characteristics for the O$_2$ plasma treatment are shown before and after annealing (measurement frequency =1Hz). (b) Comparison of measured and simulated $C$-$V$ curves for the untreated substrate with $N_A = 7.2 \times 10^{15}$ cm$^{-3}$ and $V_{FB} = +0.1$ V.

**Fig. 4:** (a) Capacitance-frequency characteristics of MIS diodes for the different substrate treatments before and after annealing. The changes due to annealing are indicated by the arrow. The curves are measured at $V_{BIAS} = -10$ V. (b) Example of loss-frequency characteristics of MIS diodes for the untreated substrate before and after annealing (at $V_{BIAS} = -10$ V). Comparison between the measured and fitted curves for annealed sample, using the 1RC and the 2RC circuit ($R_{S,1RC}$ = 2413 Ω, $C_{S,1RC}$ = 2.4 nF; $R_{S,2RC}$ = 1980 Ω, $C_{S,2RC}$ = 2.6 nF, $R_{If,2RC}$ = 1498 Ω, $C_{If,2RC}$ = 20.7 nF; the lead resistance is always negligible). The loss is defined as conductance over frequency ($G/\omega$). (c) Schematic of the equivalent circuits used for the simulations.

**Fig. 5:** (a) Output characteristics ($V_G = -10$ V) and (b, c) transfer characteristics ($V_D = -20$ V) of P3HT field-effect transistors with different substrate treatments before and after thermal annealing. The black curves are the initial ones and the grey curves after annealing. The changes are marked in Fig (b) by an arrow, except for the HMDS treatment. The transfer characteristics are shown on semi-logarithmic scale (b) and as the square root of the drain current (c).

**Fig. 6:** Temperature dependence of MIS diode and FET characteristics of samples with OTS treated substrates and thermal annealing at 420 K, measured from 200 to 410 K in steps of 10 K: (a) *C-f* characteristics in accumulation ($V_{BIAS} = -10$ V), (b) transfer characteristics in saturation ($V_D = -20$ V).

**Fig. 7:** (a) Temperature dependence of the field-effect mobility and the mobility from the bulk relaxation time in MIS diodes for different substrate treatments. The plot vs. $1/T^2$ is in accordance with the correlated disorder model. (b) Comparison of FET and MIS diode mobility for the OTS treated substrates with the predictions of an analytical model for carrier density dependent hopping transport. The used parameters are: width of the DOS $\sigma = 98$ meV, intersite spacing $a = 1.4$ nm, carrier densities $p_{MIS} = 5 \times 10^{15}$ cm$^{-3}$, $p_{FET} = 5 \times 10^{19}$ cm$^{-3}$.

**Fig. 8:** Field-effect mobility vs. water contact angle for different surface treatments of the SiO$_2$ insulator. The data shown as open circles are taken from Ref. 5.

**Fig. 1**

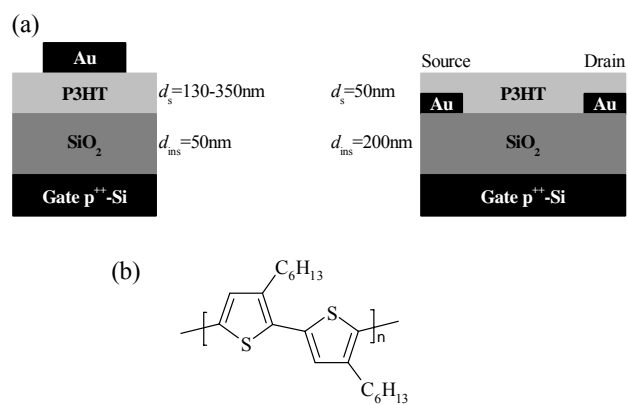

**Fig. 2**

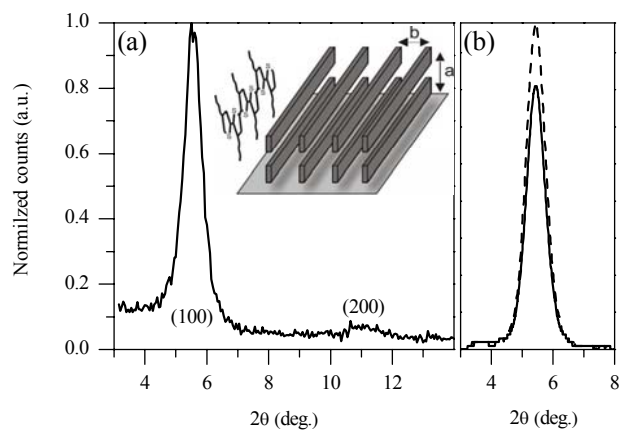

**Fig. 3**

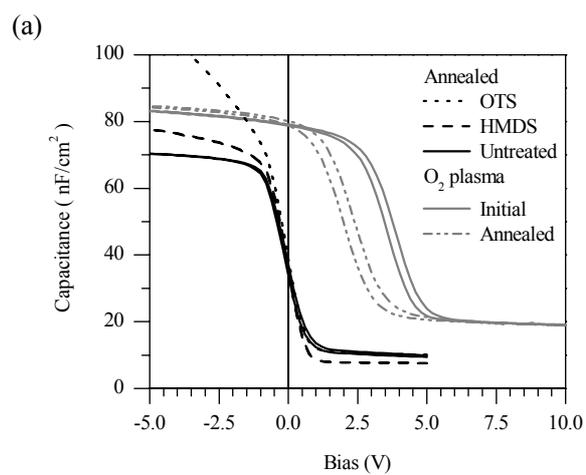
(a)

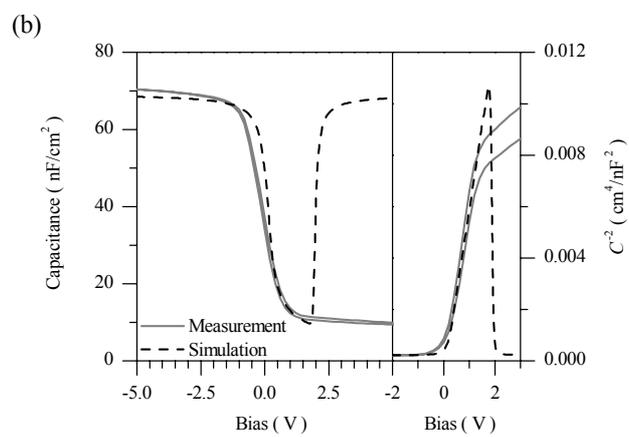
(b)

**Fig. 4**

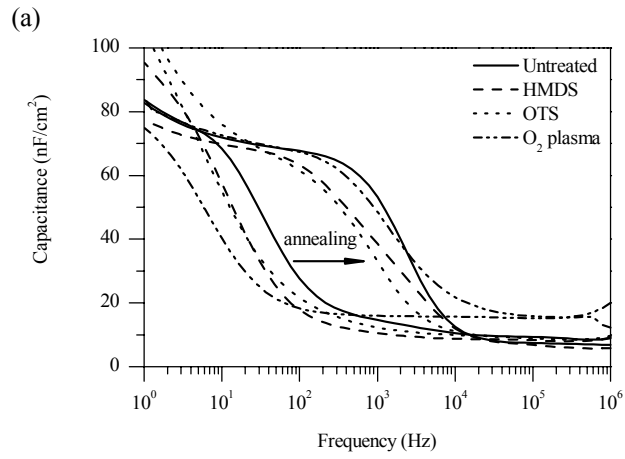

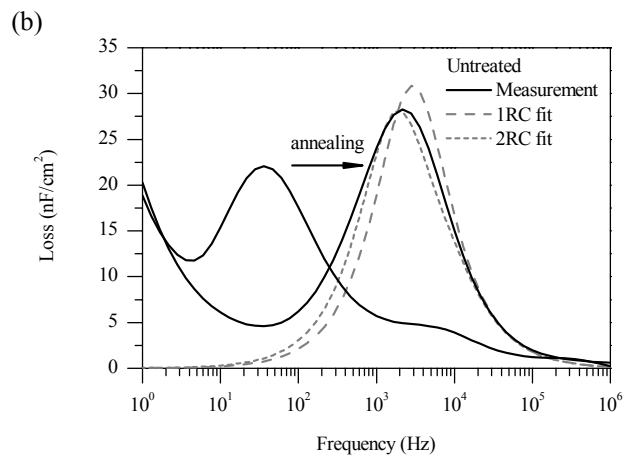

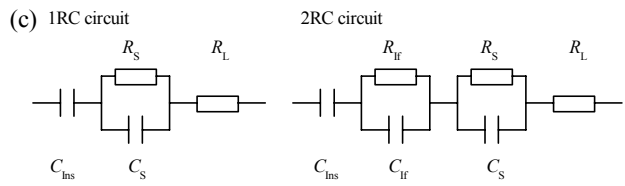

**Fig. 5**

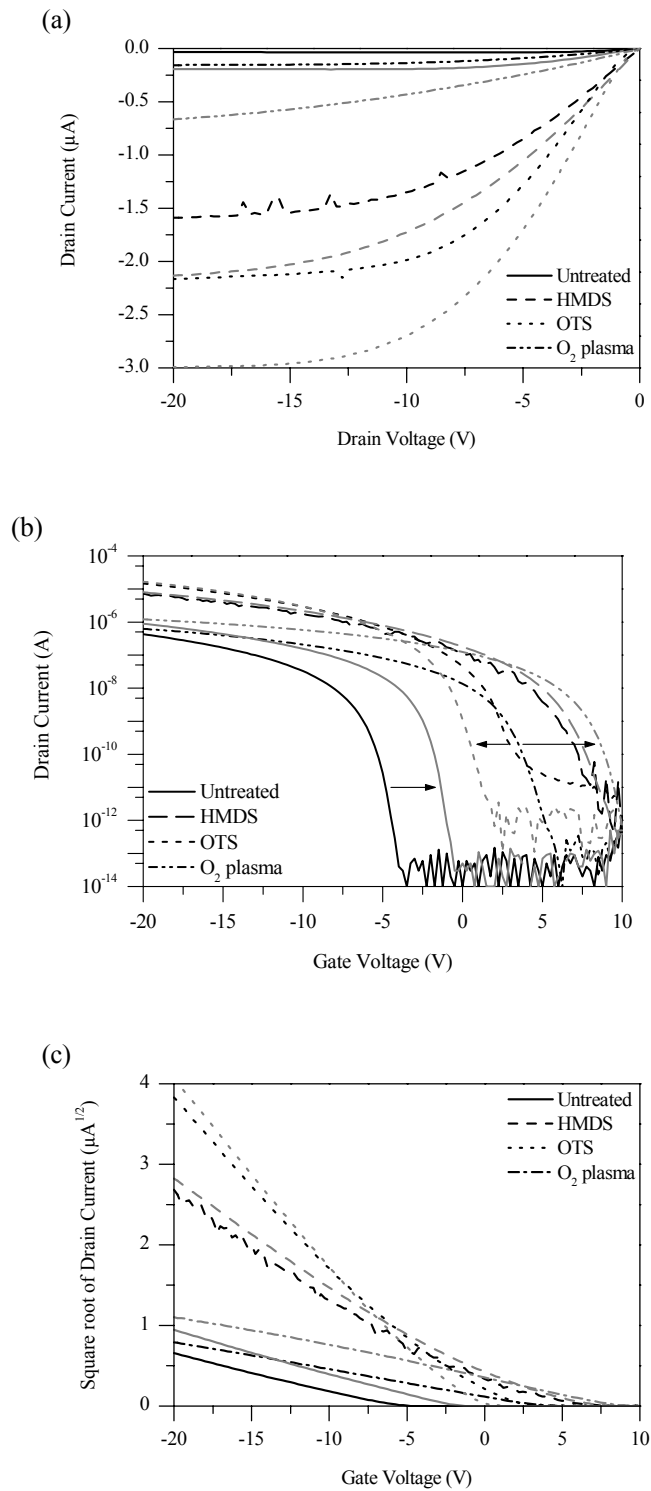

**Fig. 6**

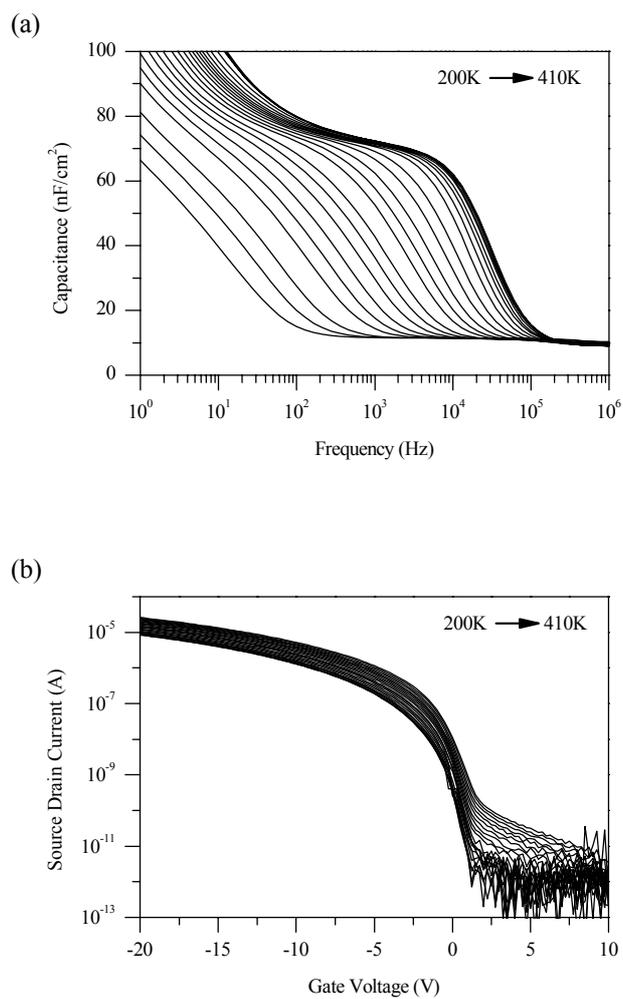

**Fig. 7**

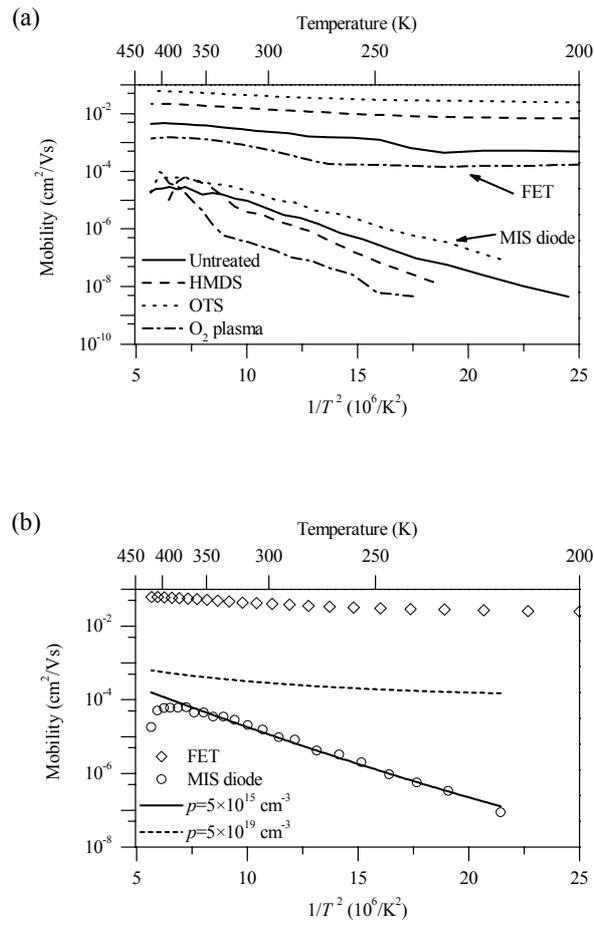

**Fig. 8**

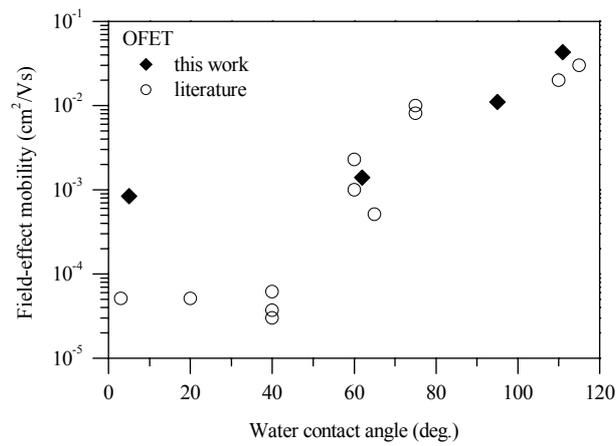

# Table captions

**Table 1:** Contact angles of water for the different types of substrates.

**Table 2:** MIS diode parameters (relaxation frequency $f_R$, doping concentration $N_A$, flatband voltage $V_{FB}$, mobility perpendicular to the interface $\mu_\perp$) for different substrate treatments and thermal annealing.

**Table 3:** OFET parameters (mobility $\mu_{sat}$, switch-on voltage $V_{SO}$, threshold voltage $V_T$, ON/OFF ratio and inverse subthreshold slope $S$) for different surface treatments before and after thermal annealing.

## Table 1

| Sample Type | $O_2$ plasma | Untreated | HMDS | OTS |
|---|---|---|---|---|
| Contact Angle(°) | <5 | 60±2 | 95±1 | 110±1 |

## Table 2

| Treatment | | $f_R$ [Hz] | $N_A$ [cm$^{-3}$] | $V_{FB}$ [V] | $\mu_\perp$ [cm$^2$/Vs] |
|---|---|---|---|---|---|
| Untreated | As prepared | 35 | 3.0×10$^{15}$ | -0.1 | 9.3×10$^{-7}$ |
| | Annealed | 2190 | 7.2×10$^{15}$ | +0.1 | 2.5×10$^{-5}$ |
| HMDS | As prepared | 10 | 3.1×10$^{15}$ | +0.25 | 2.4×10$^{-7}$ |
| | Annealed | 1135 | 2.5×10$^{15}$ | +0.2 | 3.3×10$^{-5}$ |
| OTS | As prepared | 7 | 7.1×10$^{15}$ | +0.2 | 7.8×10$^{-8}$ |
| | Annealed | 667 | 5.8×10$^{15}$ | +0.1 | 8.9×10$^{-6}$ |
| $O_2$ plasma | As prepared | 6 | 2.4×10$^{16}$ | +3.6 | 1.1×10$^{-8}$ |
| | Annealed | 1488 | 3.0×10$^{16}$ | +2.1 | 2.3×10$^{-6}$ |

**Table 3**

| Treatment | | $\mu_{sat}$ (cm$^2$/Vs) | $V_T$ (V) | $V_{SO}$ (V) | ON/OFF | $S$ (V/dec) |
|---|---|---|---|---|---|---|
| Untreated | As prepared | 1.7×10$^{-3}$ | -6.34 | -4.06 | 1.2×10$^7$ | 0.44 |
| | After annealing | 2.2×10$^{-3}$ | -2.90 | -0.65 | 2.6×10$^7$ | 0.36 |
| HMDS | As prepared | 1.3×10$^{-2}$ | -0.20 | 7.50 | 7.8×10$^7$ | 1.56 |
| | After annealing | 1.3×10$^{-2}$ | 0.90 | 9.50 | 1.9×10$^7$ | 0.85 |
| OTS | As prepared | 3.4×10$^{-2}$ | -2.30 | 3.65 | 1.4×10$^6$ | 0.92 |
| | After annealing | 4.1×10$^{-2}$ | -2.75 | 1.52 | 1.7×10$^7$ | 0.62 |
| O$_2$ plasma | As prepared | 8.4×10$^{-4}$ | 3.81 | 6.25 | 3×10$^6$ | 0.60 |
| | After annealing | 8.7×10$^{-4}$ | 12.64 | 10 | 4×10$^7$ | 0.49 |